\documentclass[reprint,showpacs,preprintnumbers,amsmath,amssymb,aps,pra]{revtex4-1}

\usepackage{graphicx}
\usepackage{dcolumn}
\usepackage{bm}

\begin{document}

\preprint{APS/123-QED}

\title{Enhanced Two-Photon Absorption in a Hollow-Core Photonic Bandgap Fiber}

\author{Kasturi Saha}%
\email{ks652@cornell.edu}
\author{Vivek Venkataraman}
\author{Pablo Londero}
\author{Alexander L. Gaeta}%
\email{a.gaeta@cornell.edu} \affiliation{School of Applied and Engineering Physics, Cornell University, Ithaca, NY-14853, USA.}

\date{\today}

\begin{abstract}
We show that two-photon absorption (TPA) in Rubidium atoms can be greatly enhanced by the use of a hollow-core photonic bandgap fiber. We investigate off-resonant, degenerate Doppler-free TPA on the $5S_{1/2}\rightarrow5D_{5/2}$ transition and observe 1$\%$ absorption of a pump beam with a total power of only 1 mW in the fiber. These results are verified by measuring the amount of emitted blue fluorescence and are consistent with the theoretical predictions which indicate that transit time effects play an important role in determining the two-photon absorption cross-section in a confined geometry.
\end{abstract}

\maketitle

\section{\label{sec:intro}Introduction:}
Two-photon absorption (TPA) is the simultaneous absorption of either two photons from a single beam of light or two single photons from two beams which results in a resonant transition from the ground state to an excited state. A key feature of TPA is that it allows access to electronic states that are otherwise dipole-forbidden single-photon transitions. It also represents one of the simplest nonlinear processes that allows one to characterize the strength of the light-matter interaction and to generate novel non-classical effects. As a result, TPA processes have also been widely studied for various nonlinear and quantum optical phenomena with applications in precision spectroscopy~\cite{Fortson}, the generation of single photons~\cite{Franson1}, the measurement of coherence and photon statistics~\cite{Osborn}, and all-optical switching~\cite{Franson2,Yavuz}. 

For many of the aforementioned applications, a significant amount of TPA is desired at low power levels. Alkali atoms have large two-photon cross-sections due to near-resonant enhancement from the intermediate state and the high oscillator strengths of the transitions involved~\cite{OscStrength,Steck}. Nevertheless, previous TPA experiments in bulk alkali vapor cells still required relatively high powers to generate measurable effects~\cite{Bloem}. One appoach to significantly reduce the threshold for observing TPA is to use waveguide geometries such as hollow-core photonic band-gap fibers (PBGF) filled with an alkali vapor since it provides tight confinement of the light in a single mode over extended distances enabling strong light-matter interactions and high nonlinearities at ultra-low power levels~\cite{lab2,lab,Lukin,lab5}. Recently, non-degenerate resonant TPA experiments at nanowatt power levels were performed in tapered optical fibers with ambient thermal Rubidium (Rb) vapor, in which the light-matter interaction occurs via the evanescent wave of the guided light in the fiber~\cite{Franson3}. It has been shown previously that a Rb-PBGF system can provide large optical depths (ODs) using light-induced atomic desorption (LIAD)~[\onlinecite{lab1}]. However, the tight confinement of atoms and photons give rise to additional spectroscopic features such as transit-time broadening which must be taken into account and has been studied previously in our system~[\onlinecite{lab3}]. 

In this paper we describe the experimental observation of efficient, Doppler-free TPA in a Rb-PBGF system using the $5S_{1/2}\rightarrow5D_{5/2}$ two-photon transition at 778.1 nm at milliwatt power levels. We estimate theoretically the amount of TPA for our system (both from direct absorption from the pump beam and from the emitted blue fluorescence) and show that our experimental results are in good agreement with the theoretical predictions. From our measurements we obtain an estimate of the transit-time of the Rb atoms in the PBGF. We extend our calculations to show how TPA can be further enhanced using a non-degenerate, on-resonance scheme.

\section{\label{sec:theory}Theoretical Calculations:}

\begin{figure}
\includegraphics{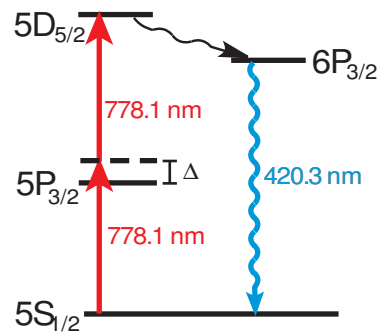}
\caption{\label{fig:Level_scheme}(Color online) Energy level structure for TPA in Rb. $\Delta$ is the detuning from the intermediate level. We observe degenerate TPA at 778.1 nm.}
\end{figure}

An illustration of the two-photon level scheme we use is shown in Fig.~\ref{fig:Level_scheme}. Once excited to the $5D_{5/2}$ level, the Rb atoms decay with 35\% probability to the intermediate $6P_{3/2}$ level, and then with 31\% probability back to $5S_{1/2}$ ground state emitting blue photons at 420.3 nm (see Fig.~\ref{fig:Level_scheme})~[\onlinecite{Branchratio}]. When the pump beam (at 778.1 nm) is far detuned from the intermediate level, or when $\Delta \gg \Gamma$, where $\Gamma$ is the linewidth of the intermediate level, then the process is termed off-resonant two-photon excitation. We characterize the nonlinearity of the Rb-PBGF system for this process by calculating the imaginary part of the third-order susceptibility $\chi^{(3)}_{Im}$~\cite{Boyd} corresponding to the relevant two-photon transition using,  
\begin{equation}
\label{eq:chi3}
\chi^{(3)}_{Im} = \frac{N \vert\mu_{1}\vert^{2} \vert\mu_{2}\vert^{2}}{\varepsilon_{o} \hbar^{3} \Delta^{2} \gamma},
\end{equation}
where \textit{N} is the atomic number density, $\mu_{1}$ and $\mu_{2}$ are the dipole moments for the $5S_{1/2}\rightarrow5P_{3/2}$ and $5P_{3/2}\rightarrow5D_{5/2}$ transitions, respectively, $\Delta$ is the detuning from the $5P_{3/2}$ level, and $\gamma$ is the homogeneous decay rate of the $5D_{5/2}$ level. In the Rb-PBGF system, $\gamma$ is determined by the transit time of the atoms across the micron-scale core of the fiber, which is much smaller than the excited state lifetime~[\onlinecite{lab3}].  

The TPA coefficient is given by
\begin{equation}
\label{eq:beta}
\beta = 0.0282 \times 10^{8} \times \left( \frac{2\pi \chi^{(3)}_{Im}}{\lambda}\right), 
\end{equation}
where $\lambda$ is the wavelength of the pump in cm. For our system, using an OD $\sim$ 100 ($ N \approx 2\times10^{13}$ atoms/$cm^{3}$), a beam waist area, $A$ $\sim$ 10$^{-7}$ cm$^{2}$, $\gamma =$ 50 MHz~[\onlinecite{lab3}], we estimate the values of $\chi^{(3)}_{Im}$ to be $4\times10^{-10}$ esu and $\beta$ to be $1.3\times10^{-6}$ cm$/$W. Correspondingly, the two-photon scattering cross-section $\sigma^{(2)}$ is estimated to be $6.5\times10^{-20}$ cm$^{4}/$W.

The transmission of the beam through a medium with a second-order absorption process can be expressed as~\cite{Boyd},
\begin{equation}
\label{eq:Iout}
T = \frac{I_{out}}{I_{in}} = \frac{1}{\left( 1 + \beta I_{in} L\right)}, 
\end{equation}
where $I_{in}$ and $I_{out}$ are the input and output intensities of the pump beam respectively, and $L$ is the interaction length. For $\beta I_{in} L\ll 1$, the percentage of absorption is given by 
\begin{equation}
\label{eq:abs}
\left[ \frac{I_{in} - I_{out}}{I_{in}} = \beta I_{in} L\right] \times 100.
\end{equation}
To estimate the fluorescence power of the emitted blue light at $\lambda_{2} = $ 420.3 nm, we take into account the branching ratios of the transitions involved during the decay from the excited $5D_{5/2}$ level to the ground $5S_{1/2}$ level, through the intermediate $6P_{3/2}$ level as discussed above. There are two possible decay mechanisms involved in the process. The first is due to a radiative decay from the $5D_{5/2}$ level, which has a lifetime $\tau_{2}$ of $\sim$240 ns, and the second is due to a non-radiative decay resulting from collisions with the walls of the fiber with a lifetime (transit-time,$\tau_{1}$) of 6 ns. Effectively, the fraction of atoms that fluoresce is $\propto \tau_{1}/(\tau_{2}+ \tau_{1}) \approx \tau_{1}/\tau_{2}$~\cite{Demtroder,Shimoda}. In addition, the decay path from $5D_{5/2} \rightarrow 5S_{1/2}$ (wavelength, $\lambda_{1} =$ 389.05 nm) involves the intermediate level $6P_{3/2}$. Hence we must also take into account the energy lost in the $5D_{5/2} \rightarrow 6P_{3/2}$ transition. Therefore, the total power of the blue fluorescence can be calculated from,
\begin{equation}
\label{eq:Bluepower}
P = (I_{in} - I_{out})\times A \times R_{1} \times R_{2} \times \frac{\tau_{1}}{\tau_{2}} \times \frac{\lambda_{1}}{\lambda_{2}},
\end{equation}
where $R_{1}(= 0.35)$ and $R_{2}(= 0.31)$ are the branching ratios of the $5D_{5/2} \rightarrow 6P_{3/2}$ and $6P_{3/2} \rightarrow 5S_{1/2}$ transitions, respectively. Using Eqs.~(\ref{eq:abs}) and ~(\ref{eq:Bluepower}) we obtain, 
\begin{equation}
\label{eq:Bluepow}
P = Const \times I_{in}^{2}.
\end{equation}
For 1-mW of power we expect to see 1$\%$ absorption in a Rb-PBGF system for a 1-cm interaction length, which is relatively significant for the degenerate off-resonant TPA process at such low pump powers. 

We next consider TPA with counter-propagating beams at 778.1 nm such that TPA occurs by the simultaneous absorption of one photon from each of the beams. The advantage of using a counter-propagating scheme is that the interaction is Doppler free~\cite{Bloem,Dopfree1,Dopfree2}. Therefore, we expect to see only homogeneously broadened lines due to transit-time effects. The following section provides the details of the experimental setup that is used to measure the blue fluorescence and direct TPA from the pump beam.  

\begin{figure*}
\includegraphics{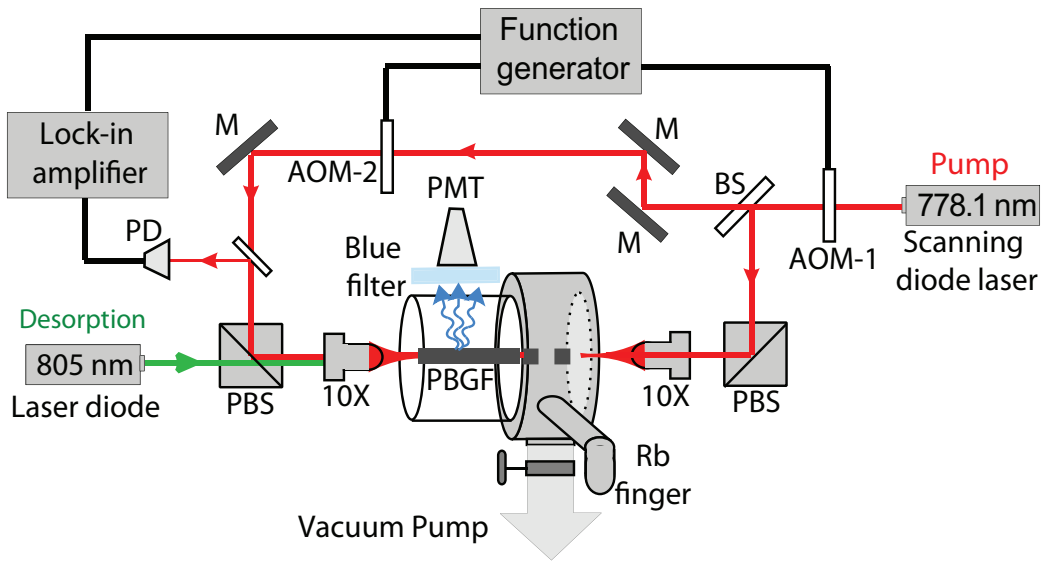}
\caption{\label{fig:Exp_Setup}(Color online) Experimental setup: The pump laser is split into two counter-propagating beams using a 50/50 beamsplitter (BS). The polarizations are made identical using polarization beam splitter (PBS) cubes, and the beams are then coupled through the hollow-core photonic bandgap fiber (PBGF). A blue colored glass is used to filter the fluorescence from the excited $6P_{3/2}$ state and then detected on a photomultiplier tube (PMT). An acousto-optic modulator (AOM-1) is used modulate the pump beams as a triangular wave at a 1-kHz frequency to vary linearly the intensity of the counter-propagating beams simultaneously. For lock-in detection of the TPA through the PBGF, one of the counter-propagating beams is modulated using AOM-2 as a square wave at 25-kHz. The signal is detected on a photodiode (PD) using a pick-off beam at one end and sent to a lock-in amplifier, the ouput of which is subsequently monitored using an oscilloscope.}
\end{figure*}

\section{\label{sec:setup}Experimental setup:}

We modified the design of the previously used Rb-PBGF system~[\onlinecite{lab1}] to allow us to image the fibers from the top and collect the emitted blue fluorecence. On  one side of the ultra high-vacuum (UHV) compatible stainless steel chamber, we attached a cylindrical glass tube that provides optical access to all sides of the fiber. A schematic of the experimental setup is presented in Fig.~\ref{fig:Exp_Setup}. We use a 9-cm-long and 6-$\mu$m-diameter hollow-core PBGF (Crystal Fiber, AIR-6-800) that guides light over the wavelength range of 750-810 nm. The fiber is mounted in a holder designed to hold multiple fibers. The entire fiber mount assembly is placed inside the vacuum chamber with an attached Rb source containing the two naturally occuring isotopes $^{85}$Rb $(\sim72\%)$ and $^{87}$Rb $(\sim28\%)$. The temperatures of the chamber and the cold-finger (Rb source) are maintained at 85$^{\circ}$C and 55$^{\circ}$C, respectively and the background pressure after bake-out is $10^{-8}$ torr. The beam from an external cavity diode laser at 778.1 nm, scanning mode-hop free across 5 GHz (over the two-photon resonances), is split into two counter-propagating beams using a 50/50 beam splitter (BS) and are each coupled to opposite ends of the fiber using 10x objectives. The two beams have equal powers and identical polarizations. 

To generate the desired OD ($\sim 100$), a highly off-resonant 20-mW desorption beam at 805 nm is also coupled into the fiber with a polarization orthogonal with respect to the pump beams. The design of our chamber enables us to collect a part of the emitted blue photons from the top. The blue photons are filtered using a blue colored glass filter and detected by a photomultiplier tube (PMT, Hamamatsu H7422P-40). The signal is then monitored and recorded for 1 second using an oscilloscope.

We measure the TPA as a function of the intensity of light field using an acouto-optic modulator (AOM-1) (see Fig.~\ref{fig:Exp_Setup}) to vary the intensity of the two-counter propagating beams simultaneously using a triangular wave at 1-kHz. The emitted blue photons are collected and detected from the top of the fiber by the method described before. For direct measurement of TPA from the 778.1 nm pump beams, we perform a lock-in detection by using another acousto-optic modulator (AOM-2) to modulate one of the two counter-propagating beams. A weak reflection of the foward propagating beam is collected from one end using a photodiode, and it's output is then sent to a lock-in amplifier, and the resultant output is then recorded using an oscilloscope.

\section{\label{sec:results}Results and Discussion:}

\subsection{\label{sec:TPAr1}Measurement of Two-Photon Absorption using Fluorescence:}

\begin{figure}
\includegraphics{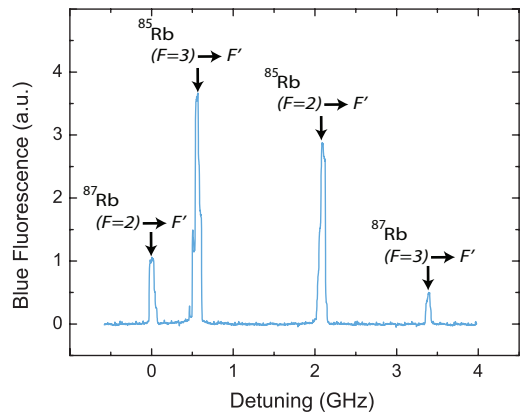}
\caption{\label{fig:Blue_Fluor}(Color online) Blue fluorescence signal detected by the PMT. Fluorescence peaks corresponding to each of the hyperfine ground states of $^{85}$Rb and $^{87}$Rb are observed as the pump laser is scanned in frequency. Since the two beams are perfectly counter-propagating in the fiber, all the peaks are Doppler-free. The peaks are homogeneously broadened due to the short transit time ($\sim$5 ns) of the Rb atoms across the $6-\mu$m core of the fiber.} 
\end{figure}

\begin{figure}
\includegraphics{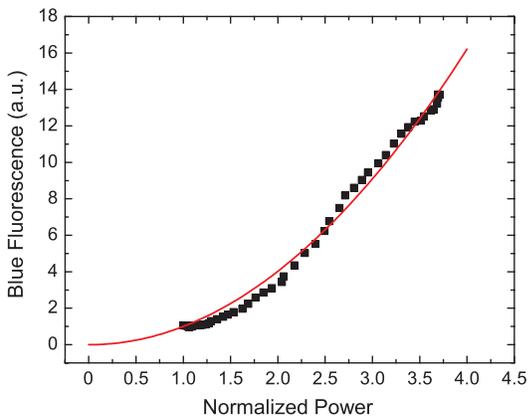}
\caption{\label{fig:Quad_Fluor_1}(Color online) Blue fluorescence signal (black dots) detected by the PMT as the pump laser power is varied using AOM-1 when the laser is tuned to the $5S_{1/2}\rightarrow5D_{5/2}$ $(F=3$ to $F^{\prime})$ two-photon transition of $^{85}$Rb. The solid red line shows the expected square dependence of two-photon absorption on intensity.} 
\end{figure}

Figure~\ref{fig:Blue_Fluor} shows the blue fluorescence signal detected by the PMT due to TPA. The four peaks correspond to each of the two hyperfine ground states of $^{85}$Rb and $^{87}$Rb. We recorded the wavelength of each of the lines using a wavemeter (Burleigh WA-1600) with an uncertainty of $\pm$0.0001 nm in each reading. For the states $F = 3$ and $F = 2$ of $^{85}$Rb, the wavelengths are 778.1055 nm and 778.1025 nm respectively, where as 778.1067 nm and 778.0997 nm correspond to the $F = 2$ and $F = 1$ states of $^{87}$Rb, respectively. The signal is Doppler free, and the line-broadening is due only to the transit-time effect. As discussed in Sec.~\ref{sec:theory}, the relatively short transit time of the Rb atoms in comparison to the lifetime of the excited $5D_{5/2}$ level suppresses the number of blue photons emitted, and we estimate the total amount of blue fluorescence emitted from the fiber taking into account the collection area of the PMT and assuming the fluorescence to be isotropic over the $4\pi$ solid angle.  We obtain 30 nW of blue light for 1 mW of pump power, which corresponds to $\sim1\%$ absorption [using Eq.~(\ref{eq:Bluepower}) and Eq.~(\ref{eq:abs})]. This agrees very well with the theoretically calculated value in Sec.~\ref{sec:theory}. Figure~\ref{fig:Quad_Fluor_1} shows the measured blue fluorescence versus input pump power, in which the power of the two counter-propagating pump beams are varied from 100 $\mu$W to 500 $\mu$W using AOM-1. We observe that the blue fluorescence increases quadratically with the intensity of the pump beams, as expected for TPA. We experimentally determine the two-photon cross-section, $\sigma^{(2)}$ to be $5\times10^{-20}$ cm$^{4}/$W for the $^{85}$Rb $5S_{1/2}\rightarrow5D_{5/2}$ transition, which agrees with the theoretically calculated value taking into account the short transit-time of the Rb atoms as the dominant decay rate.

\subsection{\label{sec:TPAr2}Direct Measurement of Two-Photon Absorption:}

\begin{figure}
\includegraphics{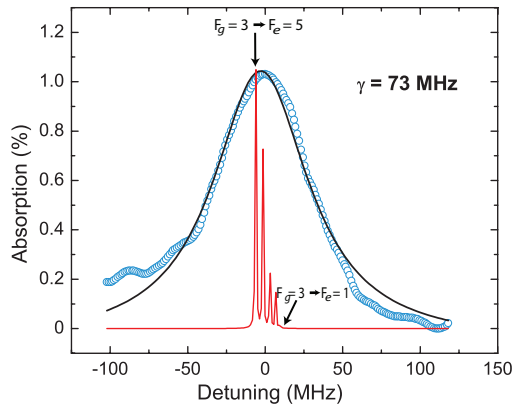}
\caption{\label{fig:Lock3}(Color online) The blue circles show the data from the direct measurement of two-photon absorption from one of the pump beams using a lock-in detector as the pump laser is scanned across the $5S_{1/2}\rightarrow5D_{5/2}$, $F_{g}=3$ to $F_{e}$ (778.1055 nm) two-photon transition of $^{85}$Rb. The solid black line shows a fit of a sum of five Lorentzians which correspond to each of the hyperfine lines. The homogenous linewidth is estimated to be 73 $\pm$10 MHz.} 
\end{figure}

\begin{figure}
\includegraphics{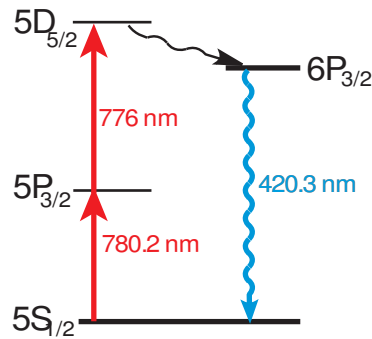}
\caption{\label{fig:Level_scheme3}(Color online) On-resonance non-degenerate two-photon absorption energy level scheme in Rb.}
\end{figure}

Due to the enhanced TPA in Rb-PBGF system, we can directly measure and quantify the amount of absorption from the pump beams. As discussed in Sec.~\ref{sec:setup}, we use a lock-in detection method since the signal-to-noise ratio in our system is low. We use AOM-2 driven by a square wave at 25 kHz to modulate one of the counter-propagating beams. A pick-off beam from the other beam is detected using a photodiode and the output is sent to a lock-in amplifier. Figure~\ref{fig:Lock3} shows the result of our measurement, and with an input pump power of 1 mW and an OD $\sim$ 100, we observe 1$\%$ absorption. This is in good agreement with our estimate of the amount of TPA from the experimental data of blue fluorescence measurement and with our theoretical estimation of TPA.

The absorption lines can also be analyzed to determine the homogeneous linewidth. We fitted a sum of five Lorentzians corresponding to the hyperfine lines ($F_{g}=3$ to $F_{e}=1,2,3,4,5$) of the $5S_{1/2}\rightarrow5D_{5/2}$ transition. The estimated linewidth is $73 \pm 10$ MHz, and represents a direct measurement of the homogeneous linewidth of Rb confined to a hollow-core photonic bandgap fiber. We also note that the homogeneous linewidth (which is determined by transit-time broadening) is slightly different from system (Rb-PBGF) to system~[\onlinecite{lab3}]. This might be due to the fact that desorption (which is used to generate the Rb vapor) occurs at a slightly different power levels and hence the corresponding temperatures and therefore the thermal velocities of the Rb vapor are slightly different. Nevertheless, the amount of off-resonance two-photon absorption is substantially high for such low pump powers, which demonstrates the potential of the Rb-PBGF system for exploring novel classical and quantum nonlinear effects at low powers. If we perform an on-resonance non-degenerate TPA~[\onlinecite{Franson3}], using the level scheme shown in Fig.~\ref{fig:Level_scheme3}, with counter-propagating 780.2 nm and 776 nm beams, we theoretically expect to see a $10^{8}$ enhancement over the off-resonance TPA. This is due to the fact that in Eq.~\ref{eq:chi3}, $\Delta = 0$; hence $\chi^{(3)}_{Im}$ is given by
\begin{equation}
\label{eq:chi3modified}
\chi^{(3)}_{Im} = \frac{N \vert\mu_{1}\vert^{2} \vert\mu_{2}\vert^{2}}{\varepsilon_{o} \hbar^{3} \gamma^{3}}.
\end{equation}
For $\gamma$ = 50 MHz, $\chi^{(3)}_{Im} = 0.18$ esu and $\sigma^{(2)}= 2.9 \times 10^{-11}$ cm$^{4}/$W. This implies that we can observe significant TPA at low photon numbers using a non-degenerate on-resonance scheme, although issues such as linear absorption must be taken into account.

\section{\label{sec:conc}Conclusion:}

In summary, we have demonstrated efficient Doppler-free degenerate TPA in Rb vapor confined to a PBGF. We observed 1$\%$ TPA with 1 mW incident pump power, and these results can readily improved by several orders of magnitude by perfoming the on-resonance TPA, which will allow for two-photon effects at sub-nanowatt input power levels. We have also been able to measure the transit-time broadening in a more direct manner. Moreover, our system enables us to explore a wide range of ODs and pump intensities in a controllable manner~[\onlinecite{lab4}]. Accessing TPA resonances in this system shows promise for exploring quantum nonlinear effects at ultra-low powers such as single-photon all optical switching, generation and measurement of non-classical states of light without the usual associated issues of inhomogeneous broadening, linear absorption and spontaneous emission noise. 

\begin{acknowledgments}
We gratefully acknowledge financial support from National Science Foundation (NSF, Grant PHY-0969996) and Defense Advanced Research Projects Agency (DARPA) under the Slow-Light program. We would also like to acknowledge useful discussions with Amar Bhagwat and Aaron Slepkov.
\end{acknowledgments}



\end{document}